\documentstyle[11pt,twoside,jltp,graphicx]{article}

\title{Electronic checkerboard pattern in striped racetrack domains: a
  consistent picture of recent neutron and STM experiments.}

\author{Brian M\o ller Andersen and Per Hedeg\aa rd}
\address{\O rsted Laboratory, Niels Bohr Institute for APG,\\
Universitetsparken 5, DK-2100 Copenhagen \O, Denmark.}

\begin{document}

\begin{abstract}
We discuss recent elastic neutron scattering and scanning tunneling
experiments on high-T$_c$ cuprates exposed to an applied magnetic field. In
particular we show that a physical picture
consisting of antiferromagnetic vortex cores operating as pinning
centers for surrounding stripes is qualitatively consistent with the
neutron data provided the stripes have the usual antiphase
modulation. Further,
we calculate the electronic structure in such a region using a T-matrix
method, and find a checkerboard interference pattern consistent with
recent scanning tunneling experiments.

\end{abstract}

\maketitle

\vspace{0.3in}

\noindent Elastic neutron scattering results on
La$_{2-x}$Sr$_{x}$CuO$_{2}$ (x=0.10) have shown that the intensity
of the incommensurate peaks in the superconducting phase is
considerably increased when a magnetic field of $H=14.5T$ is
applied perpendicular to the CuO$_{2}$ planes\cite{bella}.
Spatially resolved nuclear magnetic resonance (NMR) experiments
have shown strong evidence for antiferromagnetism in and around
the vortex cores of near-optimally doped
YBa$_2$Cu$_3$O$_{7-\delta}$\cite{mitrovic1} and
Tl$_2$Ba$_2$CuO$_{6+\delta}$\cite{kakuyanagi}. Furthermore, muon
spin rotation measurements from the mixed state of
YBa$_2$Cu$_3$O$_{6.50}$ find static antiferromagnetism in the
cores\cite{miller}. Further evidence for coexistence of the two
orders in and around vortex cores has come from scanning tunneling
microscopy (STM) performed on the surface of
YBa$_2$Cu$_3$O$_{7-\delta}$ and
Bi$_2$Sr$_2$CaCu$_2$O$_{8+x}$\cite{renner,pan,hoffmann}.\\
Theoretically the discussion of competing order parameters in the
doped Mott insulators has been a hot topic for over a
decade.
The SO(5) model was first to predict the existence of
antiferromagnetic vortex cores\cite{arovas} and to suggest
several experiments to observe these anomalous cores\cite{hallundbaek,andersen}.
\begin{figure}
\centerline{\includegraphics[height=1.0in,width=5.0in]{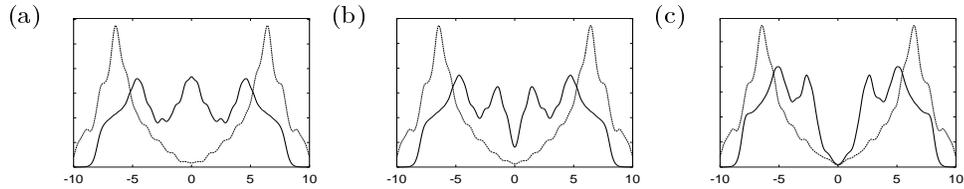}}
\caption{LDOS in the vortex core; (a) for a pure d-wave
  superconducor, (b) including a magnetic order parameter that
  increases when approaching the core (applies to YBCO), (c) same as
  (b) but with increased magnetism in the core (BSCCO). The
  dashed line shows the bulk spectrum.}
\end{figure}
Relaxing the strict SO(5) constraint
between the antiferromagnetic and the superconducting sectors seem to
be necessary to explain the large magnetic
correlation observed in ref.1\cite{hu} and to explain the modest splitting
of the zero bias conductance peak (ZBCP) expected in vortex cores of d-wave
superconductors\cite{wang,andersen2,zhu} (See figure
1). The splitting of the ZECP shown in Figure 1 is identical to the
spin splitting of the zero energy Andreev bound states at a \{110\}
interface of an antiferromagnet and a d$_{x^2-y^2}$-wave
superconductor\cite{andersen3}.\\
The physical picture we have in mind is presented in figure 2a.
\begin{figure}
\centerline{\includegraphics[height=2.0in,width=5.0in]{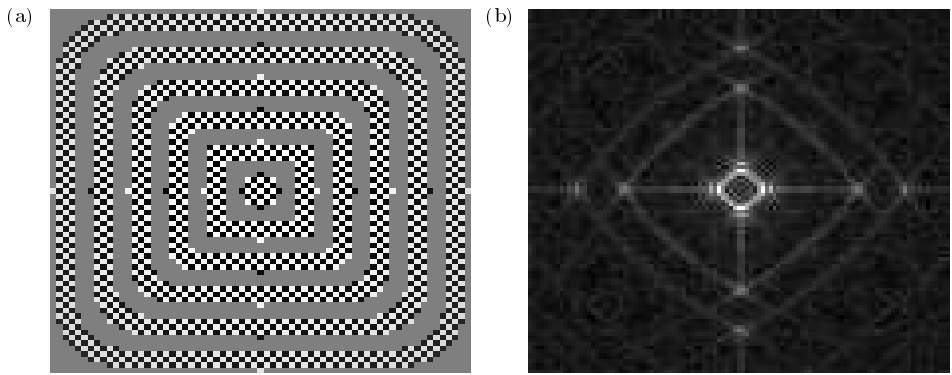}}
\caption{ a) The induced spin structure around the vortex cores. Black (white) represent spin
  up (down) and gray is the superconducting state. For clarity we have
  exaggerated the distance between adjacent magnetic domains.
  b) Fourier spectrum of the spin density order from a).}
\end{figure}
This is the ideal static picture of a real
space version of an antiferromagnetic core (center) which has pinned a
number of surrounding stripes. Without the applied magnetic field only impurities can produce a
similar pinning effect of the fluctuating stripes.
\begin{figure}
\centerline{\includegraphics[height=2.0in,width=5.0in]{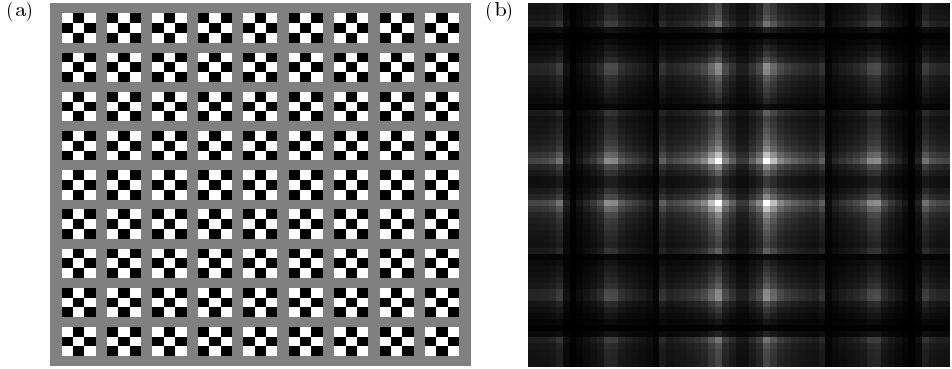}}
\caption{a) Real space picture of the spin structure in a checkerboard
  geometry. Black (white) represent spin up (down) while gray is
  the superconducting state. Similarly to the ring structure
  above each island of antiferromagnetic spins can be seen to be out
  of phase with its nearest neighbor.
  b) Fourier spectrum of the checkerboard structure shown
  in a). Note that the four main incommensurate peaks are rotated 45
  degrees relative to the Cu-O a-b axis.}
\end{figure}
In addition to the
creation of more pinning centers when applying a magnetic field, the
single site impurities are expected to pin much weaker than the large
``impurities'' created by the flux lines.
This is qualitatively
consistent with the measurements by Lake {\sl et al.}\cite{bella} of
the temperature dependence of the increased magnetic signal for
different magnetic field strengths (Note figure 2 and 3 in ref. 1).
The intention of the first part of this article is not to apply a specific model to
explain the above experiments, but to use simple Fourier analysis
to make a number of model independent observations.
In our discussion
we focus only on the competition between the ordered states and
consider complications arising from coupling to low-lying nodal
quasiparticles to be absent because of constraints from momentum
conservation.\\
Both experimentally\cite{tranquada} and
theoretically\cite{zachar} we expect an antiphase
modulation of the induced antiferromagnetic ring domains.
Indeed as seen from figure 2b the related diffraction pattern is
qualitatively consistent with the increased incommensurability
observed in ref. 1!\\
Recently STM results have shown a checkerboard pattern of
the local density of states (LDOS) around the vortex cores in slightly
overdoped BSCCO\cite{hoffmann}. Similar results obtained by Howald
{\sl et.al.}\cite{kapitulnik} without an applied magnetic field have
been succesfully described in terms of interference from quasiparticle impurity
scattering\cite{dhlee}. The LDOS modulation was found to have
half the period of the spin density wave observed by neutron
scattering (i.e. four lattice sites), and to be oriented along the
crystal axes of the CuO$_2$ plane. Is there a simple way to understand
this electronic structure in terms of some static field induced spin texture?
If we {\sl assume}, naively, that the checkerboard charge density wave is
intrinsic to the CuO$_2$ planes (and not a bilayer effect) where it
gives rise to a static spin density wave checkerboard pattern, what
should an elastic neutron scattering experiment
expect to find? To answer this question we need to investigate an
idealized pattern like the one shown in figure 3a.
Here we have also expected an antiphase order between the magnetic droplets.
The Fourier spectrum of the checkerboard spin configuration is shown
in figure 3b. Note the 45 degree rotation of the four main
incommensurable peaks and the expected checkerboard pattern of the
higher harmonics. This picture is clearly not appropriate for
LSCO for doping levels close to $x=0.10$. It is interesting that a
rotation of the incommensurable peaks at low dopings ($x < 0.055$, close
the insulator-superconductor phase transition) has
been observed in LSCO\cite{wakimoto}.
Unfortunately there is no simple
way to create an antiphase spin geometry without frustrating the spins
at low dopings where droplets of charge in an antiferromagnetic
background is the expected situation\cite{veilette}.
However, this might be possible in the highly overdoped regime
where the droplets have been inverted to separate magnetic
islands. In that case a 45 degree rotation of the incommensurable
peaks would be consistent with a checkerboard pattern.\\
\begin{figure}
\centerline{\includegraphics[height=2.0in,width=5.0in]{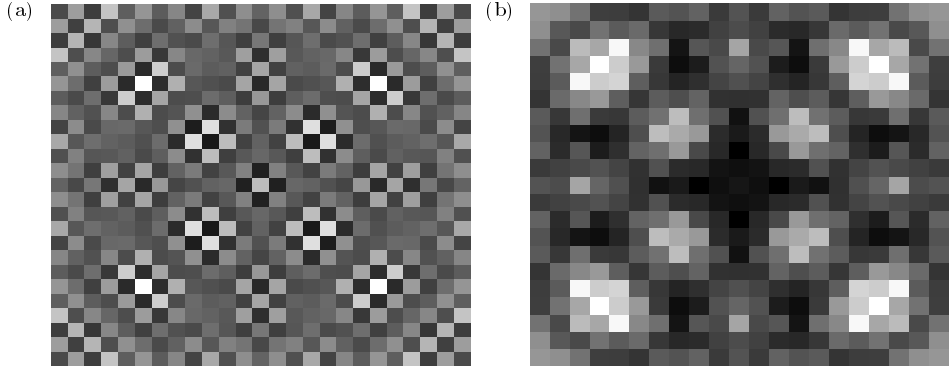}}
\caption{a) LDOS modulations in a domain of super-elliptic
  antiferromagnetic racetracks (Figure 2). b) Smeared LDOS from a)}
\end{figure}
There is, however, a simple, consistent way to understand the
racetrack spin structure of Figure 2 with the checkerboard STM
results; namely through the formation of standing waves within the
superconducting domains. Figure 4 shows the result of a full
calculation which will be presented
elsewhere\cite{andersen4}. Essentially we have embedded the
perturbations from the spin structure shown in Figure 2 in
an otherwise homogeneous d$_{x^2-y^2}$-wave superconductor and
calculated the LDOS in the perturbed region.
The figure is obtained after integrating over a small energy window
(0-12 meV) and shows clearly that checkerboard interference patterns
arise at these low energies due to interference effects. The
checkerboard LDOS pattern is very robust with respect to variations in
model parameters\cite{andersen4}.\\
In summary we discussed the phenomenology of a simple physical picture of
pinning of stripes around vortex cores that are forced
antiferromagnetic by an applied magnetic field. Contrary to commen
belief the LDOS inside the cores are not inconsistent with insulating,
magnetic cores.
The induction of striped ring structures around the core seems from
simple Fourier analysis to be consistent with the diffraction spectra
observed on LSCO only if the stripes are out of
phase with their neighbors in the usual sense.\\
In materials where a checkerboard spin pattern is relevant, we show that a 45
degree rotation of the main incommensurable peaks are to be
expected contrary to the experiments on LSCO. Finally, the
checkerboard LDOS by Hoffman {\sl et.al.} arise from
interference of standing waves around the vortices.


\begin{thebibliography}{99}
\bibitem{bella} B. Lake {\sl et.al.}, {\it Nature} {\bf 415}, 299 (2002).
\bibitem{mitrovic1} V.F. Mitrovic {\sl et.al.},
  {\it Nature} {\bf 413}, 501 (2001).
\bibitem{kakuyanagi} K. Kakuyanagi, K. Kumagai, Y. Matsuda,
  M. Hasegawa,
  {\it condmat/0206362}.
\bibitem{miller} R.I. Miller {\sl et.al.} ,
  {\it Phys. Rev. Lett}. {\bf 88}, 137002 (2002).
\bibitem{renner} I. Maggio-Aprile {\sl et.al.} Phys. Rev. Lett {\bf 75}, 2754 (1995); C. Renner,
  {\sl et.al.}, {\it Phys. Rev. Lett.} {\bf 80}, 3606 (1998).
\bibitem{pan} S.H. Pan {\sl et.al},
  {\it Phys. Rev. Lett.} {\bf 85}, 1536 (2000).
\bibitem{hoffmann} J.E. Hoffman {\sl et.al.},
  {\it Science} {\bf 295}, 466 (2002).
\bibitem{arovas} D.P. Arovas {\sl et.al.}
  {\it Phys. Rev. Lett.} {\bf 79}, 2871 (1997).
\bibitem{hallundbaek} H. Bruus {\sl et.al},
  {\it Phys. Rev. B} {\bf 59}, 4349 (1999).
\bibitem{andersen} B.M. Andersen, H. Bruus, P. Hedeg\aa rd,
  {\it Phys. Rev. B} {\bf 61}, 6298 (2000).
\bibitem{hu} J-P Hu, S-C. Zhang,
  {\it J. Phys. Chem. Solids}. (in press).
\bibitem{wang} Y. Wang, A.H. MacDonald,
  {\it Phys. Rev. B} {\bf 52}, R3876 (1995).
\bibitem{andersen2} The model calculation leading to these LDOS
  has been described elsewhere\cite{andersen}. The
  cause of the splitting is the same as that put forward by Andersen
  {\em et al.}\cite{andersen} and by Zhu {\em et
  al.}\cite{zhu}.
\bibitem{zhu} J-X. Zhu, C.S. Ting
  {\it Phys. Rev. Lett.} {\bf 87}, 147002 (2001).
\bibitem{andersen3} B.M. Andersen, P. Hedeg\aa rd,
  {\it condmat/0206422}.
\bibitem{tranquada} J.M. Tranquada {\sl et.al},
  {\it Nature} {\bf 375}, 561 (2002).
\bibitem{zachar} O. Zachar, S.A. Kivelson, V.J. Emery, {\it Phys. Rev. B}
  {\bf 57}, 1422 (1998).
\bibitem{kapitulnik} C. Howald, H. Eisaki, N. Kaneko, A. Kapitulnik,
 {\it condmat/0201546}.
\bibitem{dhlee} Q-H. Wang, D-H. Lee, {\it condmat/0205118}.
\bibitem{wakimoto} S. Wakimoto, G. Shirane, Y. Endoh, {\em et al.},
  {\it Phys. Rev. B} {\bf 60}, R769 (1999).
\bibitem{veilette} M. Veilette {\sl et.al},
  {\it Phys. Rev. Lett}. {\bf 83}, 2413 (1999).
\bibitem{andersen4} B.M. Andersen, P. Hedeg\aa rd, (unpublished).
\end{thebibliography}
\end{document}